\documentclass[aps,prl,reprint,nobibnotes,nofootinbib]{revtex4-2}

\usepackage{amsmath,amssymb,mathtools}
\usepackage{hyperref}
\hypersetup{
    colorlinks=true,
    linkcolor=blue,
    citecolor=blue,
    urlcolor=blue}

\begin{document}

\title{On Quantised Magnetic Monopoles in the Nonlocal Electromagnetic Field}

\author{E.~J.~Thompson}
\thanks{Corresponding author (Ethan Thompson)}
\email{thom3471@mylaurier.ca}
\affiliation{Transformer Lab, Toronto, Canada, M5G 0C6}
\affiliation{Wilfrid Laurier University, Waterloo, Canada, N2L 3C5}

\author{Arvin Kouroshnia}
\affiliation{Department of Physics and Astronomy, University of Waterloo, Waterloo, Ontario N2L 3G1, Canada}

\author{J. W. Moffat}
\affiliation{Perimeter Institute for Theoretical Physics, Waterloo, Ontario N2L 2Y5, Canada}
\affiliation{Department of Physics, University of Toronto, Toronto, Ontario, M5S 1A7, Canada}
\affiliation{Department of Physics and Astronomy, University of Waterloo, Waterloo, Ontario N2L 3G1, Canada}

\date{\today}

\begin{abstract}
We consider Maxwell electrodynamics in the presence of both electric and magnetic charges when the fields are made nonlocal by the prescription of the Moffat-Thompson nonlocal quantum field theory. The symmetry between the electric and magnetic equations is preserved, while a point magnetic charge is a finite distribution and its field energy becomes finite. When Dirac's quantisation condition $eg=2\pi n$ is imposed, the magnetic self-energy is found to be much greater than the corresponding electric one. We conclude that magnetic monopoles may be permitted by the theory and yet require a very large energy for their production, thus making their abundance and detection a difficult feat.
\end{abstract}

\maketitle

\emph{Introduction} -- In standard electrodynamics Maxwell's equations may become symmetric between electric and magnetic fields once magnetic charge is admitted, allowing for a magnetic monopole. In 1931 Dirac showed that the coexistence of electric and magnetic charges is constrained by the \emph{quantum condition} $eg=2\pi\;n$, meaning that the observed weak electric coupling implies a strong magnetic coupling for the smallest allowed monopole. At the same time, the field of a point monopole in local electrodynamics carries a divergent self energy or in other words; a infinite electromagnetic self-energy. 

In this paper we ask whether these two facts acquire a common interpretation in an ultraviolet dynamically nonlocal quantum theory~\cite{Moffat1990,Evens1991,MoffatThompson2026}. We show that this interpretation preserves the symmetry of the electric-magnetic duality while replacing the singular monopole core by a finite distribution width $E_M^{-1}$. The resulting magnetic field energy is finite, and Dirac quantisation then makes the minimally charged magnetic configuration parametrically more energetic than the corresponding electric one. Magnetic charge can therefore remain allowed by the fundamental equations while being exceptionally difficult to excite.

\emph{Symmetric Maxwell theory} -- In Heaviside-Lorentz units, with $c=\hbar=1$ Maxwell's equations with magnetic charges admitted may be written in the symmetric form:
\begin{align}
    \boldsymbol{\nabla}\cdot\mathbf{E} &= \rho_e,
    &
    \boldsymbol{\nabla}\cdot\mathbf{B} &= \rho_m,
    \\
    \boldsymbol{\nabla}\times\mathbf{B}
    -\frac{\partial\mathbf{E}}{\partial t}
    &=\mathbf{J}_e,
    &
    \boldsymbol{\nabla}\times\mathbf{E}
    +\frac{\partial\mathbf{B}}{\partial t}
    &=-\mathbf{J}_m ,
\end{align}
where $\rho_m$ and $\mathbf{J}_m$ denote the magnetic charge and current densities. The equations of electrodynamics then acquire a simple symmetry between the electric and magnetic sectors.  In covariant form we have that $\partial_{\mu}F^{\mu\nu}=J_e^\nu,\quad\partial_{\mu}\widetilde F^{\mu\nu}=J_m^\nu$, where $\widetilde F^{\mu\nu}=\frac{1}{2}\epsilon^{\mu\nu\rho\sigma}F_{\rho\sigma}$. The two equations are transformed into one another by an electric--magnetic rotation, provided the electric and magnetic currents are transformed at the same time. The observed world does not seem to display the same symmetry in its charged states. Electric charge is common, whereas no isolated magnetic charge has yet been established experimentally.

Dirac showed us that the coexistence of an electric charge $e$ and a magnetic charge $g$ is subject to the quantum condition $eg=2\pi n,\quad n\in\mathbb{Z}$~\cite{Dirac1931}.  For the smallest magnetic charge with $n=1$, this gives $g_D=2\pi/e$. Thus the smallness of the electric fine-structure constant, $\alpha=e^2/4\pi$, implies a large magnetic coupling, $\alpha_m=g_D^2/4\pi=1/4\alpha$. The symmetry of the field equations therefore does not require the electric and magnetic sectors to be weakly coupled at the same time.

But there is a second difficulty in the theory. A point magnetic charge in the local theory has $B_r(r)=g/4\pi r^2$, and thus an electromagnetic field energy which diverges at the origin. This divergence is usually separated from the question of charge quantization, but the two become related if the short-distance theory possesses a physical length scale.

\emph{Historical Search for Magnetic Monopoles} -- In 1894 Curie examined the physical possibility of free magnetic charge, but the subject found its modern significance when Dirac in 1931 showed that a magnetic monopole could be consistently incorporated into quantum mechanics and that the existence of a single monopole would imply the quantisation of electric charge~\cite{Curie1894,Dirac1931}. The theoretical status of monopoles changed again in 1974, when 't Hooft and Polyakov independently showed that finite-energy magnetic monopoles arise as topological soliton solutions in spontaneously broken non-Abelian gauge theories~\cite{THooft1974,Polyakov1974}. Magnetic monopoles have evolved from an optional extension of electrodynamics into objects that can emerge naturally from the global structure of gauge theories~\cite{ChoMaison1997,Kimm2015,Milton2006}.

The experimental search has followed an equally long history. Searches have been performed in cosmic radiation, terrestrial and extraterrestrial material, and high-energy accelerator collisions. A famous episode occurred in 1982, when Cabrera reported a single event in a superconducting induction detector whose persistent-current step was consistent with the passage of one Dirac magnetic charge~\cite{Cabrera1982}. No confirmed monopole population emerged from subsequent searches despite a large effort from the physics community. Large underground experiments improved the sensitivity to cosmic monopoles, for example the MACRO experiment obtained stringent flux limits over a wide interval of monopole velocities without observing a candidate~\cite{MACRO2002}. More recently, the search has moved to the LHC, where ATLAS and the dedicated MoEDAL experiment probe monopoles through their unusually large ionisation losses, while MoEDAL additionally searches for magnetic charge trapped in aluminium volumes using superconducting magnetometers~\cite{ATLAS2020,MoEDALMonopole}. But even with increasingly diverse experimental approaches, no isolated fundamental magnetic monopole has yet been discovered.

\emph{Modern Detection of Magnetic Monopoles} -- If a magnetic monopole is produced, its experimental detection need not require direct resolution of the nonlocal core as for distances $r\gg L_M$, the field obtained above approaches the ordinary Dirac monopole field and the total magnetic flux is unchanged, $\oint\mathbf{B}\cdot d\mathbf{S}=g$. A detector operating on length scales large compared with $L_M$ therefore measures the full magnetic charge. For the minimally charged Dirac monopole, $g_D/e=1/2\alpha\simeq68.5$, so that passage through matter is expected to produce an exceptionally large ionization signal~\cite{Ahlen1978}. This provides the basis for collider searches using highly ionizing tracks~\cite{ATLAS2020}. The MoEDAL experiment additionally employs trapping volumes in which a sufficiently slow monopole may become bound. The material can subsequently be passed through a superconducting quantum interference device (SQUID), where a trapped magnetic charge would produce a persistent-current signal. Searches by MoEDAL and ATLAS have so far found no established monopole candidate~\cite{MoEDAL2025,ATLAS2025}. Complementary searches for primordial or astrophysical monopoles use large-volume detectors such as IceCube, where a relativistic magnetic charge would generate an unusually intense Cherenkov signal~\cite{IceCube2022}.

Nonlocal smearing separates the energetic question of monopole production from the signatures relevant to its detection. The core is modified on scales $r\sim L_M$, but the long-distance magnetic field remains unscreened. If the nonlocal action is fundamental and there are no compensating mass contributions, its electromagnetic field energy $U_m^{\rm NL}\simeq9.66E_M$ implies a characteristic pair-production threshold $\sqrt{s}\gtrsim19.3E_M$, rather than the $13.7E_M$ scale inferred from the effective-Maxwell energy. This is a conditional kinematic estimate, not a prediction of a production cross section. Because the Dirac magnetic coupling is large, perturbative Drell--Yan and photon-fusion cross sections are not generally under perturbative control. Limits based on such benchmark mechanisms cannot be interpreted as model-independent bounds on $E_M$. Nonperturbative mechanisms, including production in the strong electromagnetic fields of ultraperipheral heavy-ion collisions, provide complementary ways to test monopole scenarios~\cite{MoEDALSchwinger2022,ATLAS2025}.

\emph{Duality-Symmetric nonlocal Electrodynamics} -- We let $\widetilde F^{\mu\nu}\equiv1/2\epsilon^{\mu\nu\rho\sigma}F_{\rho\sigma}$, and let the nonlocal operator be an entire function of the d'Alembertian with no finite zeros, $\mathcal{F}\equiv\mathcal{F}\!\left(\Box/E_M^2\right),\quad\mathcal{F}(0)=1$. We only permit the operator to be inverted and avoid the introduction of additional finite-momentum poles. In the class of nonlocal theories considered here, the same operator which modifies the ultraviolet behaviour of the propagator also defines the corresponding smearing of physical sources~\cite{ThompsonCovariance2026,ThompsonLocalization2026,ThompsonSpacetime2026}.

We take the equations of electrodynamics to be $\mathcal{F}^{-2}\partial_\mu F^{\mu\nu}=J_e^\nu$ , and $\mathcal{F}^{-2}\partial_\mu \widetilde F^{\mu\nu}=J_m^\nu$. Since $\mathcal{F}$ is invertible, these equations may equally be written in the form $\partial_\mu F^{\mu\nu}=J_{e,\mathrm{eff}}^\nu,\quad\partial_\mu \widetilde F^{\mu\nu}=J_{m,\mathrm{eff}}^\nu$, where $J_{e,\mathrm{eff}}^\nu=\mathcal{F}^{2}J_e^\nu,\quad J_{m,\mathrm{eff}}^\nu=\mathcal{F}^{2}J_m^\nu$. Electric and magnetic sources are altered by the same nonlocal map.

The symmetry is most clearly seen by introducing the complex combinations $\mathcal{G}^{\mu\nu}=F^{\mu\nu}+i\widetilde F^{\mu\nu},\quad\mathcal{J}^{\nu}=J_e^\nu+iJ_m^\nu$. The two field equations then become the single equation in the form of $\mathcal{F}^{-2}\partial_\mu\mathcal{G}^{\mu\nu}=\mathcal{J}^{\nu}$. In flat spacetime $\Box$ commutes with the dual operation, and consequently the equation is unchanged under $\mathcal{G}^{\mu\nu}\longrightarrow e^{i\theta}\mathcal{G}^{\mu\nu},\quad\mathcal{J}^{\nu}\longrightarrow e^{i\theta}\mathcal{J}^{\nu}$. Equivalently we can write this as $F^{\mu\nu}\longrightarrow F^{\mu\nu}\cos\theta+\widetilde F^{\mu\nu}\sin\theta,\quad\widetilde F^{\mu\nu}\longrightarrow\widetilde F^{\mu\nu}\cos\theta-F^{\mu\nu}\sin\theta$, with the electric and magnetic currents undergoing the same rotation. The nonlocal operator therefore does not itself distinguish electric from magnetic charge.

Current conservation is also unchanged, we see this by taking the divergence of the field equations and using the antisymmetry of $F^{\mu\nu}$ and $\widetilde F^{\mu\nu}$ gives us that $\partial_\nu J_e^\nu=0,\quad\partial_\nu J_m^\nu=0$. Ordinary Maxwell equations are recovered for wavelengths large compared with $L_M=E_M^{-1}$.

\emph{The nonlocal Magnetic Monopole} -- We now consider the field of a single magnetic charge at rest.  Let the bare magnetic density be $\rho_m(\mathbf{x})=g\,\delta^{(3)}(\mathbf{x})$, where $g$ is the total magnetic charge.  In the static limit the d'Alembertian reduces to the spatial Laplacian, and we choose the simplest Gaussian entire function,
\begin{equation}
    \mathcal{F}^{2}
    \left(\frac{\nabla^2}{E_M^2}\right)
    =
    \exp\left(\frac{\nabla^2}{E_M^2}\right).
\end{equation}
The effective source appearing in the Maxwell equation is therefore $\rho_{m,\mathrm{eff}}(\mathbf{x})=\exp\left(\nabla^2/E_M^2\right)\rho_m(\mathbf{x})$. Using the Fourier representation of the delta distribution gives us:
\begin{equation}
    \rho_{m,\mathrm{eff}}(r)
    =
    \frac{gE_M^3}{(4\pi)^{3/2}}
    \exp\left(-\frac{E_M^2r^2}{4}\right).
\end{equation}
The point magnetic charge of the local theory is replaced by a smooth distribution whose characteristic width is of order the Moffat length $L_M=E_M^{-1}$. For a static spherically symmetric configuration the nonlocal Gauss law is $\boldsymbol{\nabla}\cdot\mathbf{B}=\rho_{m,\mathrm{eff}}$. The magnetic charge contained inside a sphere of radius $r$ is consequently $g(r)=4\pi\int_0^r dr'\,r'^2\rho_{m,\mathrm{eff}}(r')$, or explicitly:
\begin{equation}
    g(r)
    =
    g\left[
    \operatorname{erf}\left(\frac{E_Mr}{2}\right)
    -
    \frac{E_Mr}{\sqrt{\pi}}
    \exp\left(-\frac{E_M^2r^2}{4}\right)
    \right].
\end{equation}
The corresponding radial field is $B_r(r)=g(r)/4\pi r^2$.

This differs essentially from the local Dirac field at short distance. Expanding for $E_Mr\ll1$, one finds that:
\begin{equation}
    g(r)
    =
    \frac{gE_M^3r^3}{6\sqrt{\pi}}
    +O(r^5),
\end{equation}
and thus $B_r(r)=gE_M^3/24\pi^{3/2}\,r
    +O(r^3)$. Thus $B_r(0)=0$, in place of the singular local behaviour $B_r\propto r^{-2}$.

At large distances the ordinary monopole field is recovered, in particular that:
\begin{equation}
    \lim_{r\rightarrow\infty}
    \oint_{S_r^2}
    \mathbf{B}\cdot d\mathbf{S}
    =
    g.
\end{equation}
The nonlocal modification therefore alters the structure of the monopole core without changing its total magnetic charge. If $g$ belongs to the Dirac charge lattice before the nonlocal map is applied, its quantized value is consequently preserved.

\emph{Action Formulation} -- The preceding field equations may also be obtained from an action principle, we first write $\mathcal{K}\equiv\mathcal{F}^{-2}\left(\Box_g/E_M^2\right)$, the gauge-field kinetic sector is described covariantly by:
\begin{equation}
    S_{\rm NL}
    =
    -\frac{1}{4}
    \int d^4x\,\sqrt{-g}\,
    F_{\mu\nu}\,
    \mathcal{K}\,
    F^{\mu\nu}
    +S_{\rm src},
\end{equation}
with the magnetic current implemented in the usual duality-symmetric two-potential completion of the source sector.  Variation of the electric and magnetic potentials then gives us that $\mathcal{K}\nabla_\mu F^{\mu\nu}=J_e^\nu,\quad\mathcal{K}\nabla_\mu\widetilde F^{\mu\nu}=J_m^\nu$ . In flat spacetime $\mathcal{K}$ is a function only of $\Box$ and commutes with $\partial_\mu$, reproducing the equations used above. Since $\mathcal{K}^{-1}=\mathcal{F}^{2}$, the same equations may again be written as ordinary Maxwell equations sourced by $J^\nu_{\rm eff}=\mathcal{F}^{2}J^\nu$. The action formulation and the effective smeared-source formulation give the same monopole field, regular core, and asymptotic magnetic charge, but the distinction becomes important only when the field energy is defined.

\emph{Effective and Action Energies} -- Although the nonlocal equation $\mathcal{K}\partial_\mu\widetilde F^{\mu\nu}=J_m^\nu$ and the effective equation $\partial_\mu\widetilde F^{\mu\nu}=\mathcal{F}^{2}J_m^\nu$ have identical classical solutions, they do not by themselves define the same energy functional. For example, if the latter is regarded as ordinary Maxwell
theory sourced by the smeared current, then:
\begin{equation}
    U_m^{\rm eff}
    =
    \frac{1}{2}\int d^3x\,\mathbf B^2 .
\end{equation}
For $\widetilde{\rho}_{m,\rm eff}(\mathbf{k})
=g\exp(-\mathbf{k}^2/E_M^2)$ this gives us:
\begin{equation}
    U_m^{\rm eff}
    =
    \frac{g^2}{2}
    \int\frac{d^3k}{(2\pi)^3}
    \frac{e^{-2\mathbf{k}^2/E_M^2}}{\mathbf{k}^2}
    =
    \frac{g^2E_M}{16\pi}
    \sqrt{\frac{2}{\pi}} .
\end{equation}
For the minimally charged Dirac monopole, $g_D^2=\pi/\alpha$, this is $U_m^{\rm eff}\simeq6.83E_M$.

The fundamental nonlocal action contains the operator $\mathcal{K}$ in its kinetic term. Its conserved field energy must therefore be obtained from the stress-energy tensor of the nonlocal action rather than by assigning the local Maxwell density $\mathbf B^2/2$ to the same field.  The two values should consequently be regarded as the effective-Maxwell and action-level energy assignments, respectively, not as different monopole solutions.

\emph{Finite Energy and the Dirac Condition} -- In the effective Maxwell representation used above this energy is:
\begin{equation}
    U_m
    =
    \frac{1}{2}
    \int d^3x\,\mathbf{B}^2 .
\end{equation}
It is simplest to evaluate the integral in momentum space. The Gaussian source has Fourier transform $\widetilde{\rho}_{m,\mathrm{eff}}(\mathbf{k})=g\exp\left(-\mathbf{k}^2/E_M^2\right)$, and Gauss' law gives us:
\begin{equation}
    \widetilde{\mathbf B}(\mathbf{k})
    =
    -i\frac{\mathbf{k}}{\mathbf{k}^2}
    \widetilde{\rho}_{m,\mathrm{eff}}(\mathbf{k}).
\end{equation}
Consequently we see that:
\begin{align}
    U_m
    &=
    \frac{1}{2}
    \int\frac{d^3k}{(2\pi)^3}
    \frac{
    \left|
    \widetilde{\rho}_{m,\mathrm{eff}}(\mathbf{k})
    \right|^2
    }{\mathbf{k}^2}
    \nonumber\\
    &=
    \frac{g^2}{2}
    \int\frac{d^3k}{(2\pi)^3}
    \frac{
    e^{-2\mathbf{k}^2/E_M^2}
    }{\mathbf{k}^2}.
\end{align}
The integral is elementary and gives $U_m=(g^2E_M/16\pi)\sqrt{2/\pi}$. The divergence of the local point monopole self-energy is replaced by a finite energy proportional to the nonlocal scale.

The same calculation for a unit electric charge gives $U_e=(e^2E_M/16\pi)\sqrt{2/\pi}$. At this stage the theory remains symmetric under the interchange of electric and magnetic charge.  The distinction enters through the quantum condition on the allowed charges. For the smallest Dirac monopole $eg_D=2\pi$, so that $g_D^2=4\pi^2/e^2=\pi/\alpha,\quad e^2=4\pi\alpha$. The magnetic energy is therefore $U_m=E_M/16\alpha\sqrt{2/\pi}$ for the minimally charged monopole, whereas for the electric field $U_e=\alpha E_M/4\sqrt{2/\pi}$. Their ratio is $U_m/U_e=1/4\alpha^2$. Taking $\alpha^{-1}\simeq137$, this is $U_m/U_e\simeq4.7\times10^3$. Numerically we have $U_m\simeq6.8\,E_M,\quad U_e\simeq1.46\times10^{-3}E_M$.

This large difference has not been put into the nonlocal equations. The same form factor, the same length scale, and the same field equations have been used for the two charges. It follows instead from Dirac's condition that describes the weakness of the observed electric coupling requires the elementary magnetic charge to be large. The effective-Maxwell result is an energy assignment to the smeared field, not necessarily the conserved electromagnetic contribution of the underlying nonlocal action or the complete physical monopole mass.

The quantity $U_m$ should be understood as the electromagnetic contribution to the energy of the monopole, and not by itself as a complete determination of its physical mass. A microscopic monopole may possess further core or matter contributions. I the absence of a compensating negative contribution, the result establishes a characteristic energetic scale of order $M_m \gtrsim O\!\left(E_M/\alpha\right)$, and thus suggests that magnetic charge can be allowed by a duality-symmetric theory while remaining costly to produce.

\emph{Stress-Energy from the Nonlocal Action} -- To determine the action-level energy, we define the Hilbert stress-energy tensor by:
\begin{equation}
    T_{\mu\nu}
    =
    -\frac{2}{\sqrt{-g}}
    \frac{\delta S_{\rm NL}}{\delta g^{\mu\nu}} .
\end{equation}
The metric dependence of the entire operator must be included. If we consider $\mathcal{K}(\Box_g)=\sum_{n=0}^{\infty}c_n\Box_g^n$, its variation is:
\begin{equation}
    \delta\mathcal{K}
    =
    \sum_{n=1}^{\infty}c_n
    \sum_{\ell=0}^{n-1}
    \Box_g^\ell
    (\delta\Box_g)
    \Box_g^{\,n-1-\ell}.
\end{equation}
For the stationary monopole, $F_{0i}=0$ and $\partial_0F_{\mu\nu}=0$.  After integration by parts the terms generated by $\delta\Box_g$ enter $T^{00}$ as spatial improvement terms, so that in flat spacetime $T^{00}=\frac{1}{2}\mathbf B\cdot\mathcal{K}\mathbf B+\partial_i\Xi^i$ , where the explicit form of $\Xi^i$ is not required for the total conserved energy. The monopole satisfies $\mathbf B=O(r)$ at the origin and $\mathbf B=O(r^{-2})$ asymptotically, so the corresponding surface term vanishes. Hence:
\begin{equation}
    U_m^{\rm NL}
    =
    \int d^3x\,T^{00}
    =
    \frac{1}{2}
    \int d^3x\,
    \mathbf B\cdot\mathcal{K}\mathbf B .
\end{equation}
For the Gaussian form factor used above:
\begin{equation}
    \mathcal{K}(\mathbf{k})
    =
    e^{\mathbf{k}^2/E_M^2},
    \qquad
    \widetilde{\mathbf B}(\mathbf{k})
    =
    -ig\frac{\mathbf{k}}{\mathbf{k}^2}
    e^{-\mathbf{k}^2/E_M^2},
\end{equation}
and therefore:
\begin{align}
    U_m^{\rm NL}
    &=
    \frac{g^2}{2}
    \int\frac{d^3k}{(2\pi)^3}
    \frac{e^{-\mathbf{k}^2/E_M^2}}
    {\mathbf{k}^2}
    \nonumber\\
    &=
    \frac{g^2E_M}{8\pi^{3/2}} .
\end{align}
For $eg_D=2\pi$ this becomes:
\begin{equation}
    U_m^{\rm NL}
    =
    \frac{E_M}{8\alpha\sqrt{\pi}}
    \simeq
    9.66\,E_M .
\end{equation}
So inclusion of the metric variation of the nonlocal operator does not alter the integrated coefficient, but it modifies the local stress-energy density by improvement terms whose contribution vanishes after spatial integration for the regular static monopole. The factor $U_m^{\rm NL}/U_m^{\rm eff}=\sqrt{2}$ instead follows because the effective-Maxwell energy contains two Gaussian smearing factors, whereas the nonlocal kinetic operator cancels one of them. Both energies remain finite, but the action-level value is the appropriate electromagnetic contribution when the nonlocal action is taken as fundamental.

\emph{Energy Conservation and the Absence of a Thermodynamic Pathology} -- 
Now one may conclude that we could be violating the first law of thermodynamics and thus be able to build a perpetual motion machine. But there is a nuanced difference between $U_m^{\rm eff}$ and $U_m^{\rm NL}$, so then it does not imply that energy may be extracted by cycling between the two descriptions. They are not two physically available energies of the same state, but they arise from different energy functionals. If the nonlocal action is fundamental, the conserved energy of the complete system is obtained from that action, including the monopole source or core degrees of freedom. Time-translation invariance, or equivalently diffeomorphism invariance before taking the flat-space limit, implies on shell $\partial_\mu T_{\rm tot}^{\mu\nu}=0,\quad T_{\rm tot}^{\mu\nu}=T_{\rm NL}^{\mu\nu}+T_{\rm core}^{\mu\nu}+T_{\rm int}^{\mu\nu}$, then we see that:
\begin{equation}
    \frac{dE_{\rm tot}}{dt}
    =
    -\oint_{\partial V} dS_i\,T_{\rm tot}^{i0}.
\end{equation}
For an isolated system whose energy flux vanishes at spatial infinity, $dE_{\rm tot}/dt=0$, so any closed creation--annihilation cycle satisfies $\Delta E_{\rm tot}=0,\quad\oint dE_{\rm tot}=0$. So unfortunately we cannot create a monopole using $U_m^{\rm eff}\simeq6.83E_M$ and annihilate it to recover $U_m^{\rm NL}\simeq9.66E_M$. The former is the Maxwell energy assigned to the smeared field profile, whereas the latter is the conserved field contribution defined by the fundamental nonlocal action, by mixing the two amounts to using different Hamiltonians for the two halves of the cycle. The $\sqrt{2}$ difference does not furnish a violation of the first law or a perpetual-motion mechanism.

\emph{Consequences for Monopole Production} -- The preceding results gives us a simple reason why magnetic charge may be difficult to produce even when it is admitted by the fundamental field equations. The nonlocal theory does not suppress the total magnetic charge, since the asymptotic flux remains $g$, but it assigns a finite energy to the field concentrated within the region $r\lesssim L_M$. When the nonlocal action is taken as fundamental, the appropriate conserved field-energy contribution for the minimally charged Dirac monopole is:
\begin{equation}
U_m^{\rm NL}
=
\frac{E_M}{8\alpha\sqrt{\pi}}
\simeq
9.66,E_M .
\end{equation}
This should be distinguished from the effective-Maxwell assignment $U_m^{\rm eff}\simeq6.83E_M$, obtained by evaluating $1/2\int d^3x,\mathbf B^2$ for the same classical field. The two expressions correspond to the same monopole solution, but the former follows from the conserved stress-energy tensor of the fundamental nonlocal action. If the physical monopole mass is written schematically as $M_m=U_m^{\rm NL}+M_{\rm core}+\Delta M$,
where $M_{\rm core}$ represents any additional matter or topological contribution and $\Delta M$ is the quantum corrections, then in the absence of a compensating negative contribution one expects $M_m\gtrsim U_m^{\rm NL}$. The kinematic threshold for monopole--antimonopole pair production would consequently satisfy $\sqrt{s}\gtrsim2M_m\gtrsim E_M/(4\alpha\sqrt{\pi})$, where $\sqrt{s}$ is the total centre-of-mass energy. For the Gaussian realization considered here this gives the characteristic electromagnetic scale $\sqrt{s}_{\rm mag}\gtrsim19.3,E_M$, before any additional monopole-core or quantum contribution is included. This scale is a kinematic energetic requirement rather than a prediction for the monopole-production cross section, which may depend strongly on the microscopic production mechanism.

The absence of observed monopoles need not imply that electric--magnetic duality is absent from the fundamental equations. A duality-symmetric nonlocal theory can preserve magnetic charge and Dirac quantisation while making the magnetic sector energetically much more difficult to excite than the electric sector. The asymmetry arises not from an explicit breaking of the electromagnetic field equations or of the nonlocal action, but from the Dirac quantum condition relating the allowed electric and magnetic charges.

\emph{Conclusion} -- We have shown that a duality-symmetric nonlocal electrodynamics can admit magnetic charge while removing the singular core of the point monopole. The nonlocal form factor preserves the asymptotic magnetic flux and Dirac charge quantisation while replacing the singular source by a smooth distribution. The resulting electromagnetic energy is finite. Although the equivalent effective-Maxwell description gives $U_m^{\rm eff}\simeq6.83E_M$, variation of the fundamental nonlocal action gives the conserved action-level energy $U_m^{\rm NL}\simeq9.66E_M$ for the minimally charged Dirac monopole. Because the same nonlocal operator acts on both electric and magnetic sectors, their action-level energies retain the ratio ($U_m^{\rm NL}/U_e^{\rm NL}=1/(4\alpha^2)\simeq4.7\times10^3$). In the absence of compensating contributions to the monopole mass, this corresponds to a characteristic minimum electromagnetic contribution to the pair-production threshold of $\sqrt{s}\gtrsim19.3E_M$. So the absence of observed monopoles may reflect not the exclusion of magnetic charge, but the large energetic cost imposed by Dirac quantisation when the ultraviolet electromagnetic field is governed by a fundamentally nonlocal action.

\end{document}